\documentstyle[12pt,axodraw]{article}
\begin{document}
\baselineskip 0.6cm
\newcommand{\gsim}{ \mathop{}_{\textstyle \sim}^{\textstyle >} }
\newcommand{\lsim}{ \mathop{}_{\textstyle \sim}^{\textstyle <} }
\def\barr{\bar}


\begin{titlepage}

\begin{flushright}
UCB-PTH-01/35 \\
LBNL-48827 \\
\end{flushright}

\vskip 0.5cm

\begin{center}
{\Large \bf  Strongly Coupled Grand Unification \\
 in Higher Dimensions}

\vskip 1.0cm

{\large 
Yasunori Nomura
}

\vskip 0.5cm
 {\it Department of Physics, \\ and \\
Theoretical Physics Group, Lawrence Berkeley National Laboratory,\\
University of California, Berkeley, CA 94720}

\vskip 1.0cm

\abstract{
We consider the scenario where all the couplings in the theory are 
strong at the cut-off scale, in the context of higher dimensional 
grand unified field theories where the unified gauge symmetry is 
broken by an orbifold compactification.  In this scenario, the 
non-calculable correction to gauge unification from unknown 
ultraviolet physics is naturally suppressed by the large volume 
of the extra dimension, and the threshold correction is dominated 
by a calculable contribution from Kaluza-Klein towers that gives 
the values for $\sin^2\theta_w$ and $\alpha_s$ in good agreement 
with low-energy data.  The threshold correction is reliably 
estimated despite the fact that the theory is strongly coupled at 
the cut-off scale.  A realistic 5d supersymmetric $SU(5)$ model is 
presented as an example, where rapid $d=6$ proton decay is avoided 
by putting the first generation matter in the 5d bulk.}

\end{center}
\end{titlepage}


\section{Introduction}
\label{sec:intro}

The unification of the three gauge couplings around $M_U \sim 2 \times 
10^{16}~{\rm GeV}$ \cite{Georgi:1974yf} in the minimal supersymmetric 
standard model strongly suggests new physics at this energy scale.  
Conventionally, this new physics has been viewed as 4 dimensional grand 
unified theories (GUTs) \cite{Georgi:1974sy}, in which all the standard 
model gauge interactions are unified into a single non-Abelian gauge 
group and quarks and leptons are unified into smaller numbers of 
representations under the gauge group.  Grand unification in 
4 dimensions (4d), however, raises several new questions, including 
how the GUT gauge symmetry is broken, why the doublet and triplet 
components of Higgs multiplets split, and why we have not already observed 
proton decay caused by color triplet Higgsino exchanges \cite{Sakai:1982pk}.

On the other hand, these questions have also been addressed in the 
context of higher dimensional theories in string theory.  In this case, 
the grand unified group is broken by boundary conditions imposed on 
the gauge field, and the triplet Higgses are projected out from the 
zero-mode sector, leaving only the doublet Higgses as massless 
fields \cite{Candelas:1985en, Dixon:jw}.  This is possible because 
there is no zero-mode gauge symmetry which transforms massless doublet 
Higgses into massless triplet Higgses.  In this framework, however, 
there is no field theoretic unified symmetry remaining at low energy, 
so that we have to resort to string threshold calculations to tell 
whether the three gauge coupling constants are really unified at the 
string scale \cite{Ibanez:1991zv}.

Recently, we have introduced a new framework in which the gauge coupling 
unification is realized in higher dimensional unified field theories 
compactified to 4d on orbifolds \cite{Hall:2001pg}.  Kawamura first 
suggested an $SU(5)$ GUT in 5d \cite{Kawamura:2001ev}, using an 
$S^1/(Z_2 \times Z_2')$ orbifold earlier introduced in the supersymmetry 
breaking context \cite{Barbieri:2001vh}.  A completely realistic theory 
was obtained in Ref.~\cite{Hall:2001pg}, where it was shown that 
a special field theoretic symmetry called restricted gauge symmetry 
plays a crucial role in this type of theories.  This restricted gauge 
symmetry arises from the fact that there is a moderately large energy 
interval where the physics is described by higher dimensional grand 
unified {\it field} theories.  In the higher dimensional picture it has 
gauge transformation parameters whose dependence on the extra dimensional 
coordinates is constrained by orbifold boundary conditions; in the 4d 
picture it is a symmetry that has different Kaluza-Klein (KK) 
decompositions for the ``unbroken'' and ``broken'' gauge transformations.
Using these ideas, various higher dimensional GUT models have been 
constructed \cite[10 --15]{Kawamura:2001ev, Hall:2001pg}.

In the specific case of 5d $SU(5)$ models in Refs.~\cite{Kawamura:2001ev, 
Hall:2001pg}, the 5d $SU(3)_C \times SU(2)_L \times U(1)_Y$ (3-2-1) 
gauge transformation has a KK decomposition in terms of $\cos[2ny/R]$, 
while the 5d $SU(5)/(SU(3)_C \times SU(2)_L \times U(1)_Y)$ (X-Y) has 
a decomposition in terms of $\cos[(2n+1)y/R]$.  Since there is no zero 
mode for the X-Y gauge transformation, the restricted gauge symmetry 
does not require that the doublet and triplet Higgses must have the 
same mass.  However, due to higher KK tower gauge transformations, 
local operators written in the 5d bulk must still preserve the complete 
$SU(5)$ symmetry, and all the $SU(5)$-breaking local operators must be 
located on the fixed point where only 3-2-1 gauge symmetry is preserved 
\cite{Hall:2001pg}.  In particular, $SU(5)$-violating effects from 
unknown ultraviolet physics must appear as boundary operators on this 
fixed point.  This is crucial for guaranteeing the successful gauge 
coupling unification in this framework.  Since the $SU(5)$-violating 
contributions to the gauge couplings which come from the fixed point 
are suppressed by the volume of the extra dimension compared with the 
$SU(5)$-preserving contribution from the bulk, we can argue that the 
gauge coupling is (approximately) unified, without invoking any string 
theory calculation, if the volume of the extra dimension is sufficiently 
large \cite{Hall:2001pg}.  Then, small deviations from the case of 
exact unification at a single threshold scale become calculable and 
improve the agreement between the experimental value and theoretical 
prediction of $\sin^2\theta_w$ \cite{Hall:2001pg, Nomura:2001mf}.  
The gauge coupling unification in higher dimensional GUTs has been 
further studied using dimensional deconstruction \cite{Csaki:2001qm, 
Cheng:2001qp} and dimensional regularization \cite{Contino:2001si}.

In view of the important role played by the large volume for the 
successful prediction of $\sin^2\theta_w$, in this paper we study the 
possibility that the theory has the maximally large volume allowed by 
strong coupling analysis.  We consider the scenario where all the 
couplings in the theory are strong at the cut-off scale and show that 
it is consistent with observations.  In this paper we restrict our 
analysis to an order-of-magnitude level, leaving detailed numerical 
studies for future work.  We present a realistic 5d supersymmetric 
$SU(5)$ model as an explicit example.  The model preserves the 
successful $b/\tau$ Yukawa unification and does not have the unwanted 
$SU(5)$ mass relations for the first two generations.  It also partially 
explains fermion mass hierarchies due to the configuration of the matter 
fields in the extra dimension.  We find that the observed values of 
the low-energy gauge couplings are well reproduced if we take the volume 
of the extra dimension to be large as suggested by the strong coupling 
analysis.  Experimental signatures from $d=6$ proton decay are also 
discussed, and it is shown that the final state generically contains 
the second or third generation particles.  Finally, the values of 
the cut-off and compactification scales obtained by analyzing gauge 
couplings give a 4d Planck scale close to the observation, giving 
a clue of how to solve the conventional problem in string theory of 
separating the string and the apparent unification scales.

The paper is organized as follows.  In the next section, we give a 5d 
supersymmetric $SU(5)$ model that can accommodate the large volume of 
the extra dimension without conflicting with the constraint from $d=6$ 
proton decay.  In section \ref{sec:gcu}, we consider the gauge coupling 
unification in this model and argue that the model is consistent with 
low-energy data.  The $d=6$ proton decay and the 4d Planck scale are 
discussed in section \ref{sec:other}.  Finally, our conclusions are 
drawn in section \ref{sec:concl}.

\section{Minimal Model}
\label{sec:model}

In this paper, we consider a minimal realization of the scenario where 
the theory is strongly coupled at the cut-off scale.  Thus, we consider 
a single extra dimension and the smallest grand unified group, $SU(5)$.  
It should, however, be noted that we present this case as a 
representative example of more general scenario.  We begin with briefly 
reviewing the bulk structure of 5d supersymmetric $SU(5)$ theories 
\cite{Kawamura:2001ev, Hall:2001pg}.  The 5d spacetime is a 
direct product of 4d Minkowski spacetime $M^4$ and an extra dimension 
compactified on the $S^1/(Z_2 \times Z_2')$ orbifold, with coordinates 
$x^\mu$ $(\mu = 0,1,2,3)$ and $y$ $(=x^5)$, respectively.  The 
$S^1/(Z_2 \times Z_2')$ orbifold can be viewed as a circle of radius 
$R$ divided by two $Z_2$ transformations; $Z_2$: $y \to -y$ and 
$Z_2'$: $y' \to -y'$ where $y' = y - \pi R/2$.  Here, $R$ is around 
the GUT scale, $R \sim M_U^{-1}$.  The physical space is an interval 
$y: [0, \pi R/2]$ which has two branes at the two orbifold fixed 
points at $y=0$ and $\pi R/2$. 

Under the $Z_2 \times Z_2'$ symmetry, a generic 5d bulk field 
$\phi(x^\mu, y)$ has a definite transformation property
\begin{eqnarray}
  \phi(x^\mu, y) \to \phi(x^\mu, -y) &=& P \phi(x^\mu, y), \\
  \phi(x^\mu, y') \to \phi(x^\mu, -y') &=& P' \phi(x^\mu, y'),
\end{eqnarray}
where the eigenvalues of $P$ and $P'$ must be $\pm 1$.  Denoting the 
field with $(P, P') = (\pm 1, \pm 1)$ by $\phi_{\pm \pm}$, 
we obtain the following mode expansions \cite{Barbieri:2001vh}: 
\begin{eqnarray}
  \phi_{++} (x^\mu, y) &=& 
      \sum_{n=0}^{\infty} \frac{1}{\sqrt{2^{\delta_{n,0}} \pi R}} \, 
      \phi^{(2n)}_{++}(x^\mu) \cos{2ny \over R},
\label{eq:expansion-1}
\\
  \phi_{+-} (x^\mu, y) &=& 
      \sum_{n=0}^{\infty} \frac{1}{\sqrt{\pi R}} \,
      \phi^{(2n+1)}_{+-}(x^\mu) \cos{(2n+1)y \over R},
\\
  \phi_{-+} (x^\mu, y) &=& 
      \sum_{n=0}^{\infty} \frac{1}{\sqrt{\pi R}} \,
      \phi^{(2n+1)}_{-+}(x^\mu) \sin{(2n+1)y \over R},
\\
  \phi_{--} (x^\mu, y) &=& 
      \sum_{n=0}^{\infty} \frac{1}{\sqrt{\pi R}} \,
      \phi^{(2n+2)}_{--}(x^\mu) \sin{(2n+2)y \over R},
\label{eq:expansion-4}
\end{eqnarray}
where the 4d fields $\phi^{(2n)}_{++}$, $\phi^{(2n+1)}_{+-}$, 
$\phi^{(2n+1)}_{-+}$, and $\phi^{(2n+2)}_{--}$ acquire masses 
$2n/R$, $(2n+1)/R$, $(2n+1)/R$, and $(2n+2)/R$ upon compactification.
Zero modes are contained only in $\phi_{++}$ fields, so that the matter 
content of the massless sector is smaller than that of the full 
5d multiplet.

In the 5d bulk, we have $SU(5)$ gauge supermultiplets and two Higgs 
hypermultiplets that transform as ${\bf 5}$ and ${\bf 5}^*$.  The 5d 
gauge supermultiplet contains a vector boson, $A_M$ ($M=0,1,2,3,5$), 
two gauginos, $\lambda$ and $\lambda'$, and a real scalar, $\sigma$, 
which is decomposed into a vector supermultiplet, $V(A_\mu, \lambda)$, 
and a chiral multiplet in the adjoint representation, 
$\Sigma((\sigma+iA_5)/\sqrt{2}, \lambda')$, under $N=1$ supersymmetry 
in 4d.  The hypermultiplet, which consists of two complex scalars, 
$\phi$ and $\phi^c$, and two Weyl fermions, $\psi$ and $\psi^c$, 
forms two 4d $N=1$ chiral multiplets, $\Phi(\phi, \psi)$ and 
$\Phi^c(\phi^c, \psi^c)$, transforming as representations conjugate to 
each other under the gauge group.  Here $\Phi$ runs over the two Higgs 
hypermultiplets, $H$ and $\barr{H}$.  ($\{ H, \barr{H}^c \}$ and 
$\{\barr{H}, H^c \}$ transform as ${\bf 5}$ and ${\bf 5}^*$ under 
the $SU(5)$, respectively.)

\begin{table}
\begin{center}
\begin{tabular}{|c|c|c|}
\hline
 $(P,P')$ & 4d $N=1$ superfield & mass       \\ \hline
 $(+,+)$  & $V^a$, $H_F$, $\barr{H}_F$                  & $2n/R$     \\ 
 $(+,-)$  & $V^{\hat{a}}$, $H_C$, $\barr{H}_C$          & $(2n+1)/R$ \\ 
 $(-,+)$  & $\Sigma^{\hat{a}}$, $H_C^c$, $\barr{H}_C^c$ & $(2n+1)/R$ \\ 
 $(-,-)$  & $\Sigma^a$, $H_F^c$, $\barr{H}_F^c$         & $(2n+2)/R$ \\ 
\hline
\end{tabular}
\end{center}
\caption{The $(Z_2, Z_2')$ transformation properties for the bulk 
gauge and Higgs multiplets.}
\label{ta:Z2-Z2}
\end{table}
The 5d $SU(5)$ gauge symmetry is ``broken'' by the orbifold 
compactification to a 4d $SU(3)_C \times SU(2)_L \times U(1)_Y$ gauge 
symmetry by choosing $P=(+,+,+,+,+)$ and $P'=(-,-,-,+,+)$ acting on the 
${\bf 5}$ \cite{Kawamura:2001ev}.  Each $Z_2$ reflection is taken to 
preserve the same 4d $N=1$ supersymmetry.  The $(Z_2, Z_2')$ charges 
for all components of the vector and Higgs multiplets are shown in 
Table~\ref{ta:Z2-Z2}.  Here, the indices $a$ and $\hat{a}$ denote the 
unbroken and broken $SU(5)$ generators, $T^a$ and $T^{\hat{a}}$, 
respectively. The $C$ and $F$ represent the color triplet and weak 
doublet components of the Higgs multiplets, respectively: 
$H \supset \{ H_C, H_F \}$, $\barr{H} \supset \{ \barr{H}_C, \barr{H}_F \}$, 
$H^c \supset \{ H_C^c, H_F^c \}$, and $\barr{H}^c \supset 
\{ \barr{H}_C^c, \barr{H}_F^c \}$.  Since only $(+,+)$ fields have 
zero modes, the massless sector consists of $N=1$ $SU(3)_C \times 
SU(2)_L \times U(1)_Y$ vector multiplets $V^{a(0)}$ with two Higgs 
doublet chiral superfields $H_F^{(0)}$ and $\barr{H}_F^{(0)}$.  The 
higher modes for the vector multiplets $V^{a(2n)}$ $(n>0)$ eat 
$\Sigma^{a(2n)}$ becoming massive vector multiplets, and similarly 
for the $V^{\hat{a}(2n+1)}$ and $\Sigma^{\hat{a}(2n+1)}$ $(n \geq 0)$.  
Since the non-zero modes for the Higgs fields have mass terms of the 
form $H_F^{(2n)} H_F^{c(2n)}$, $\barr{H}_F^{(2n)} \barr{H}_F^{c(2n)}$, 
$H_C^{(2n+1)} H_C^{c(2n+1)}$, and $\barr{H}_C^{(2n+1)} 
\barr{H}_C^{c(2n+1)}$, there is no dimension 5 proton decay from 
color triplet Higgsino exchange \cite{Hall:2001pg}.

Now, we consider the gauge couplings in our scenario.  Here we roughly 
estimate various quantities at the tree level; more detailed discussions 
including radiative corrections are given in section \ref{sec:gcu}.
Since we require that the theory is strongly coupled at the cut-off scale 
$M_*$, the gauge kinetic terms are given by 
\begin{eqnarray}
  S = \int d^4x\, dy \int d^2\theta \left[ 
    \frac{\eta M_*}{16 \pi^3} {\cal W}^{\alpha} {\cal W}_{\alpha} 
  + \delta(y) \frac{\eta'}{16 \pi^2} {\cal W}^{\alpha} {\cal W}_{\alpha}
  + \delta(y-\frac{\pi}{2}R) \frac{\eta'_i}{16 \pi^2} 
    {\cal W}_i^{\alpha} {\cal W}_{i\alpha} \right]
  + {\rm h.c.},
\end{eqnarray}
where we have used naive dimensional analysis (NDA) in higher 
dimensions.\footnote{
In Ref.~\cite{Chacko:2000hg} a different coefficient of $M_*/24\pi^3$ 
was used for the bulk kinetic term, which was derived by considering 
loop expansions in the non-compactified 5d space.  Here we use 
$M_*/16\pi^3$ instead, since it correctly reproduces the strong-coupling 
value for the 4d gauge coupling, $g \simeq 4\pi/(M_* R)^{1/2}$, 
after integrating out the extra dimension and is more appropriate 
in the case of the compactified space.  The coefficients of 
brane-localized terms are determined by requiring that all loop expansion 
parameters are order one in the 4d picture.}
Here, $\eta$, $\eta'$, and $\eta'_i$ are order one coefficients and 
$i$ runs over $SU(3)_C$, $SU(2)_L$, and $U(1)_Y$.  The restricted gauge 
symmetry requires that the first two terms must preserve the $SU(5)$ 
symmetry.  The last term, however, can have different coefficients 
for $i= SU(3)_C, SU(2)_L, U(1)_Y$, which encode $SU(5)$-violating 
effects from unknown physics above the cut-off scale.  After integrating 
over the extra dimension, we obtain the zero-mode gauge couplings 
at the cut-off scale as
\begin{eqnarray}
  \frac{1}{g_{i}^2} 
  = \frac{\eta M_* R}{16 \pi^2} + \frac{\eta'}{16 \pi^2}
    + \frac{\eta'_i}{16 \pi^2}.
\label{eq:coupling}
\end{eqnarray}
Since we know that $1/g_i^2 \sim 1$ from the observed values of 
the low-energy gauge coupling constants, the ratio between the 
compactification and the cut-off scales must be $M_* R \sim 
16 \pi^2 / g_i^2 = O(10^2 - 10^3)$.\footnote{
The actual value of $M_* R$ could be somewhat smaller than the naive 
estimate given here, due to a group theoretical factor $C$ appearing 
in loop expansions: $M_* R \sim 16 \pi^2 / C g_i^2$.}
We find that the threshold correction from unknown ultraviolet physics 
above $M_*$ is suppressed by $1/(M_* R) \sim 1/(16 \pi^2)$ and thus 
negligible in the present scenario.  Therefore, the threshold correction 
to $\sin^2\theta_w$ is dominated by the calculable contribution coming 
from an energy interval between $1/R$ and $M_*$.

We next consider the configuration of matter fields.  Since 
$M_* R \gsim 100$ corresponds to $1/R \lsim 10^{15}~{\rm GeV}$, it 
requires that the first generation matter must live in the bulk; 
otherwise $d=6$ proton decay occurs much faster rate than experimental 
constraints allow \cite{Hall:2001pg}.\footnote{
The ${\bf 5}^*$ of the first generation may be located in the bulk 
without conflicting with the bound from proton decay.}
On the other hand, to preserve successful $b/\tau$ unification in 
supersymmetric GUTs \cite{Chanowitz:1977ye}, we have to put the third 
generation matter on the $SU(5)$-preserving brane located at $y=0$, 
since if we put quarks and leptons in the bulk there are no $SU(5)$ 
Yukawa relations \cite{Hall:2001pg}.\footnote{
Models without the $b/\tau$ unification are obtained if we put the 
third generation ${\bf 5}^*$ in the bulk.}
These considerations almost fix the location of the matter fields.  
The remaining choices are only concerning where we put ${\bf 10}$ and 
${\bf 5}^*$ of the second generation.  Since we do not want the $SU(5)$ 
relation, $m_s = m_\mu$, for the second generation, at least one of 
${\bf 10}$ and ${\bf 5}^*$ must be put in the bulk.  Thus, we are left 
with three possibilities: (i) both ${\bf 10}$ and ${\bf 5}^*$ 
in the bulk, (ii) ${\bf 10}$ in the bulk and ${\bf 5}^*$ on the 
$y=0$ brane, (iii) ${\bf 10}$ on the $y=0$ brane and ${\bf 5}^*$ 
in the bulk.  As we will see later, the second possibility may be 
preferred in view of quark and lepton mass matrices, especially in 
view of the large mixing angle between the second and third generation 
neutrinos observed in the Super-Kamiokande experiment \cite{Fukuda:1998mi}.  
We therefore take this possibility as an illustrative purpose for the 
moment.  We consider all three possibilities when we discuss $d=6$ 
proton decay later.\footnote{
In these modes, supersymmetry breaking may occur through the mechanism 
of Ref.~\cite{Barbieri:2001yz} that uses small parameters appearing in 
boundary conditions.  In this case, the first possibility of both 
${\bf 10}$ and ${\bf 5}^*$ in the bulk is preferred to suppress 
flavor violating contributions to the first-two generation sfermion 
masses.  The flavor violation would then occur in the processes involving 
the third generation particles.  One way of avoiding all these concerns 
is to consider 6d models in which gaugino mediation \cite{Kaplan:2000ac} 
works while suppressing $d=6$ proton decay \cite{Hall:2001zb}.
We leave detailed phenomenologies of these models including supersymmetry 
breaking for future work.}

We now explicitly present our model.  The gauge and Higgs sectors are 
as discussed before.  For the matter fields, we introduce the third 
generation matter chiral superfields $T_3({\bf 10})$, $F_3({\bf 5}^*)$ 
and the second generation one $F_2({\bf 5}^*)$ on the $y=0$ brane.
In the 5d bulk, we have to introduce six hypermultiplets 
${\cal T}_2 = \{ T_2({\bf 10}), T_2^c({\bf 10}^*) \}$, 
${\cal T}'_2 = \{ T'_2({\bf 10}), T_2^{\prime c}({\bf 10}^*) \}$, 
${\cal T}_1 = \{ T_1({\bf 10}), T_1^c({\bf 10}^*) \}$, 
${\cal T}'_1 = \{ T'_1({\bf 10}), T_1^{\prime c}({\bf 10}^*) \}$, 
${\cal F}_1 = \{ F_1({\bf 5}^*), F_1^c({\bf 5}) \}$, 
${\cal F}'_1 = \{ F'_1({\bf 5}^*), F_1^{\prime c}({\bf 5}) \}$, 
to obtain the correct low-energy matter content.
The transformations for these bulk matter fields under 
$Z_2 \times Z_2'$ are given by $P=(+,+,+,+,+)$ and $P'=(-,-,-,+,+)$ 
acting on the ${\bf 5}$ for unprimed fields, but for primed fields 
the $Z_2'$ quantum numbers are assigned to be the opposite of the 
corresponding unprimed fields \cite{Hall:2001pg} (for details, see 
Refs.~\cite{Hebecker:2001wq, Barbieri:2001yz}).  Then, the quark 
and lepton zero modes come from various brane and bulk fields as
\begin{eqnarray}
&& T_3  \supset Q_3, U_3, E_3, \qquad
   F_3  \supset D_3, L_3, 
\label{eq:3rd-gen}\\
&& T_2  \supset U_2, E_2, \qquad
   T'_2 \supset Q_2, \qquad
   F_2  \supset D_2, L_2,
\label{eq:2nd-gen}\\
&& T_1  \supset U_1, E_1, \qquad
   T'_1 \supset Q_1, \qquad
   F_1  \supset L_1, \qquad
   F'_1  \supset D_1.
\label{eq:1st-gen}
\end{eqnarray}
Since the first generation quarks and leptons which would be unified 
into a single multiplet in the usual 4d GUTs come from different 
$SU(5)$ multiplets, proton decay from broken gauge boson exchange is 
absent at the leading order \cite{Hall:2001pg}.  (This result is 
also obtained from KK momentum conservation in the fifth dimension.)

The Yukawa couplings are written on the $y=0$ brane.  On this brane, 
all the operators of the form $[T T H]_{\theta^2}$ and 
$[T F \barr{H}]_{\theta^2}$ are written with the size of their 
coefficients dictated by NDA in higher dimensions.  Here, $T$ and $F$ runs 
over $\{ T_3, T_2, T'_2, T_1, T'_1 \}$ and $\{ F_3, F_2, F_1, F'_1 \}$, 
respectively.  Similar Yukawa couplings can also be written at $y=\pi R/2$ 
brane for matter in the bulk.  After integrating over $y$, we obtain 
the Yukawa matrices for low-energy quarks and leptons.  At the 
compactification scale, they take the form
\begin{eqnarray}
  {\cal L}_4 &\simeq& 
    \sqrt{\frac{16 \pi^2}{M_* R}} 
  \pmatrix{
     {\bf 10}_1 & {\bf 10}_2 & {\bf 10}_3 \cr
  }
  \pmatrix{
     \epsilon^2 & \epsilon^2 & \epsilon \cr
     \epsilon^2 & \epsilon^2 & \epsilon \cr
     \epsilon   & \epsilon   & 1        \cr
  }
  \pmatrix{
     {\bf 10}_1 \cr {\bf 10}_2 \cr {\bf 10}_3 \cr
  } H
\nonumber\\
  && + 
    \sqrt{\frac{16 \pi^2}{M_* R}} 
  \pmatrix{
     {\bf 10}_1 & {\bf 10}_2 & {\bf 10}_3 \cr
  }
  \pmatrix{
     \epsilon^2 & \epsilon & \epsilon \cr
     \epsilon^2 & \epsilon & \epsilon \cr
     \epsilon   & 1        & 1        \cr
  }
  \pmatrix{
     {\bf 5}^*_1 \cr {\bf 5}^*_2 \cr {\bf 5}^*_3 \cr
  } \barr{H},
\end{eqnarray}
where $\epsilon \simeq (M_* R)^{-1/2} \sim 0.1$ and we have omitted 
order-one coefficients.  Low-energy quark and lepton fields are defined 
as ${\bf 10}_i \equiv \{ Q_i, U_i, E_i \}$ and 
${\bf 5}^*_i \equiv \{ D_i, L_i \}$ $(i = 1,2,3)$, 
so that they generically contain fields coming from different 
hypermultiplets (see Eqs.~(\ref{eq:3rd-gen} -- \ref{eq:1st-gen})).
Here, we have normalized these fields canonically in 4d.  In the above 
equation, the matrix elements denoted as $\epsilon$ or $\epsilon^2$ 
do not respect $SU(5)$ relations, while the ones denoted as $1$ must 
respect $SU(5)$ relations since they entirely come from the Yukawa 
couplings among the matter fields localized on the $SU(5)$-preserving 
($y=0$) brane.  Therefore, the model does not have unwanted $SU(5)$ 
fermion mass relations for the first two generations, while preserving 
the $b/\tau$ unification \cite{Hall:2001zb}.

Since $\sqrt{16 \pi^2/M_* R} \sim g_i$, the present model predicts 
$y_t \sim g_i$ at the compactification scale, which is in reasonably 
good agreement with low-energy data.  The over-all mass difference 
between up- and down-type quarks should be given by 
$\tan\beta \equiv \langle H_F \rangle / \langle \barr{H}_F \rangle 
\sim 50$.\footnote{
In the case of ${\bf 5}^*_3$ in the bulk, we obtain 
$\tan\beta \sim (m_t/m_b)\epsilon \sim 5$.}
This large value of $\tan\beta$ may also be compatible with the 
$b/\tau$ Yukawa unification \cite{Pierce:1997zz}.  The above mass 
matrices roughly explain the observed pattern of quark and lepton 
masses and mixings; for example, the presence of the mass hierarchy 
between the first-two generation and the third generation fermions.  
To reproduce the detailed structure of fermion masses in the first two 
generations, however, there must be some cancellations among different 
elements and/or small numbers in coefficients of order 
$10^{-1} - 10^{-2}$.  It will be interesting to look for the model 
where more complicated structure gives completely realistic fermion 
mass matrices \cite{Hall:2001rz}.

How about neutrino masses?  Small neutrino masses are obtained by 
introducing right-handed neutrino fields $N$ through the see-saw 
mechanism \cite{Seesaw}.  They can be introduced either on the $y=0$ 
brane or in the 5d bulk, and have Yukawa couplings of the form 
$[F N H]_{\theta^2}$ and Majorana masses of the form $[N N]_{\theta^2}$
at the $y=0$ brane.  After integrating out $N$ fields, we obtain the mass 
matrix for the light neutrinos of the form 
\begin{eqnarray}
  {\cal L}_4 &\simeq& 
    \frac{1}{M_R} \left( \frac{16 \pi^2}{M_* R} \right) 
  \pmatrix{
     {\bf 5}^*_1 & {\bf 5}^*_2 & {\bf 5}^*_3 \cr
  }
  \pmatrix{
     \epsilon^2 & \epsilon & \epsilon \cr
     \epsilon   & 1        & 1        \cr
     \epsilon   & 1        & 1        \cr
  }
  \pmatrix{
     {\bf 5}^*_1 \cr {\bf 5}^*_2 \cr {\bf 5}^*_3 \cr
  } H H,
\end{eqnarray}
regardless of the configuration of the right-handed neutrino fields.
The over-all mass scale $M_R$ is given by right-handed neutrino Majorana 
masses, which we here assume to be provided by some other physics such 
as $U(1)_{B-L}$ breaking scale.  An interesting point is that the present 
matter configuration naturally explains the observed large mixing angle 
between the second and third generation neutrinos, by putting both the 
second and third generation ${\bf 5}^*$'s on the $y=0$ brane.

\section{Gauge Coupling Unification}
\label{sec:gcu}

In this section, we show that the observed values of the low-energy 
gauge couplings are well reproduced if the volume of the extra dimension 
is large as is suggested by the strong coupling analysis.  We also argue 
that the situation in the present scenario is better than in usual 
4d GUTs, since the masses for the GUT-scale particles are completely 
determined by KK mode expansions.

Let us first estimate the radiative corrections to the gauge couplings 
coming from loops of KK towers whose masses lie between $1/R$ and 
$M_*$.  In the 4d picture, the zero-mode gauge couplings $g_i$ at the 
compactification scale $M_c$ ($=1/R$) are given by
\begin{eqnarray}
  \frac{1}{g_i^2(M_c)} \simeq 
    \frac{1}{g_0^2(M_*)} - \frac{b}{8 \pi^2} (M_* R - 1)
    + \frac{b'_i}{8 \pi^2} \ln(M_* R),
\label{eq:rge-5d}
\end{eqnarray}
where $b$ and $b'_i$ are constants of $O(1)$.  The second and third 
terms on the right-hand side represent the pieces which run by power-law 
and logarithmically.  A crucial observation made in Refs.~\cite{Hall:2001pg, 
Nomura:2001mf} is that the coefficient $b$ is necessarily $SU(5)$ 
symmetric, since the power-law contributions come from renormalizations 
of 5d kinetic terms which must be $SU(5)$ symmetric due to the restricted 
gauge symmetry.  The logarithmic contributions come from renormalizations 
of 4d kinetic terms localized on the branes, and can be different for 
$SU(3)_C$, $SU(2)_L$, and $U(1)_Y$.  Thus, gauge coupling unification 
is logarithmic even above the compactification scale.  This situation 
is quite different from the power-law unification scenario of 
Ref.~\cite{Dienes:1998vh}.

Since the power-law piece is asymptotically non-free in the 
present set-up, the ratio between the compactification and 
cut-off scales could be smaller than the purely classical estimate.  
This power-law contribution also has a sensitivity to the ultraviolet 
physics.  However, it is expected that this does not change the order 
of magnitude of the tree-level estimate of $M_* R$, since the theory 
is strongly coupled only around the cut-off scale and is weakly coupled 
over a wide energy range from $1/R$ to $M_*$.  Therefore, we here take 
$M_* R \simeq 100$ as a representative value.  Note that we have 
ambiguities coming from $\eta$'s in Eq.~(\ref{eq:coupling}) in any case, 
so that the precise value is not very important at this stage.

To calculate the effect of the KK towers on the gauge coupling 
unification, we consider the one-loop renormalization group equations 
for the three gauge couplings \cite{Hall:2001pg, Nomura:2001mf}.  
Since the $SU(5)$-violating contribution to the gauge couplings from 
unknown ultraviolet physics above $M_*$ is suppressed by the large 
volume, we set the three gauge couplings equal to a unified value 
$g_*$ at $M_*$.  Then, the equations take the following form:
\begin{eqnarray}
  \alpha_i^{-1}(m_Z) &=& \alpha_*^{-1}(M_*)
    + \frac{1}{2\pi} \Biggl\{ a_i \ln\frac{m_{\rm SUSY}}{m_Z} 
    + b_i \ln\frac{M_*}{m_Z} \nonumber\\
  && \qquad \qquad 
    + c_i \sum_{n=0}^{N_l} \ln\frac{M_*}{(2n+2)M_c}
    + d_i \sum_{n=0}^{N_l} \ln\frac{M_*}{(2n+1)M_c} \Biggr\}, 
\label{eq:rge}
\end{eqnarray}
where $(a_1, a_2, a_3) = (-5/2, -25/6, -4)$, $(b_1, b_2, b_3) = 
(33/5, 1, -3)$, $(c_1, c_2, c_3) = (6/5+n_{{\bf 5}^*}+3n_{\bf 10}, 
-2+n_{{\bf 5}^*}+3n_{\bf 10}, -6+n_{{\bf 5}^*}+3n_{\bf 10})$, and 
$(d_1, d_2, d_3) = (-46/5+n_{{\bf 5}^*}+3n_{\bf 10}, 
-6+n_{{\bf 5}^*}+3n_{\bf 10}, -2+n_{{\bf 5}^*}+3n_{\bf 10})$.
Here, we have assumed a common mass $m_{\rm SUSY}$ for the superparticles 
for simplicity, and the sum on $n$ includes all KK modes below $M_*$, 
so that $(2N_l+2)M_c \leq M_*$; $n_{{\bf 5}^*}$ and $n_{\bf 10}$ 
represent the numbers of generations which are put in the bulk 
($n_{{\bf 5}^*} = 1$ and $n_{\bf 10} = 2$ in the present case).  Taking 
a linear combination of the three equations, we obtain
\begin{eqnarray}
  (5 \alpha_1^{-1} - 3 \alpha_2^{-1} - 2 \alpha_3^{-1})(m_Z) 
  = \frac{1}{2\pi} \Biggl\{ 8 \ln\frac{m_{\rm SUSY}}{m_Z} 
  + 36 \ln\frac{(2N_l+2)M_c}{m_Z} 
  - 24 \sum_{n=0}^{N_l} \ln\frac{(2n+2)}{(2n+1)} \Biggr\},
\end{eqnarray}
where we have set $M_* = (2N_l+2)M_c$.  Note that $n_{{\bf 5}^*}$ and 
$n_{\bf 10}$ drop out from this equation, since a combination of bulk 
hypermultiplets whose massless modes give a complete $SU(5)$ 
representation has $SU(5)$ symmetric matter content at each KK mass 
level.  Since the corresponding linear combination in the usual 4d 
minimal supersymmetric $SU(5)$ GUT takes the form
\begin{eqnarray}
  (5 \alpha_1^{-1} - 3 \alpha_2^{-1} - 2 \alpha_3^{-1})(m_Z) 
  = \frac{1}{2\pi} \Biggl\{ 8 \ln\frac{m_{\rm SUSY}}{m_Z} 
  + 36 \ln\frac{M_U}{m_Z} \Biggr\},
\end{eqnarray}
where $M_U = (M_\Sigma^2 M_V)^{1/3}$ \cite{Hisano:1992mh}, we find 
the following correspondence between the two theories:
\begin{eqnarray}
  \ln\frac{M_c}{m_Z} =
    \ln\frac{M_U}{m_Z} 
    + \frac{2}{3} \sum_{n=0}^{N_l} \ln\frac{(2n+2)}{(2n+1)}
    - \ln(2N_l+2),
\label{eq:correspond-1}
\end{eqnarray}
as far as the running of the gauge couplings is concerned.  An important 
point here is that this KK contribution improves the agreement between 
the experimental value and theoretical prediction of $\sin^2\theta_w$ 
and $\alpha_s$ \cite{Hall:2001pg, Nomura:2001mf}.  This is because 
$b'_i$'s in Eq.~(\ref{eq:rge-5d}) are given by $b'_i = b_i - c_i/2$, 
and are not equal to the low-energy $\beta$-function coefficients, $b_i$, 
plus some universal pieces.  At the leading order, the contributions 
from KK towers to $\sin^2\theta_w$ and $\alpha_s$ are given by 
$\Delta_{\sin^2\theta_w} \simeq -(1/5\pi) \alpha \ln(M_* R)$ and 
$\Delta_{\alpha_s} \simeq -(3/7\pi) \alpha_s^2 \ln(M_* R)$, respectively, 
which well reproduce experimental values with $M_* R \simeq 10^2 - 10^3$.
A more detailed analysis including the next to leading order effect 
has been given in Ref.~\cite{Contino:2001si}, where it was shown that 
if $M_* R \simeq 100$ the KK contribution would indeed give the right 
values for $\sin^2\theta_w$ and $\alpha_s$ in a reasonable range 
of $m_{\rm SUSY}$.

Using the experimental values of the gauge couplings, we obtain 
$1 \times 10^{16}~{\rm GeV} \lsim M_U \lsim 3 \times 10^{16}~{\rm GeV}$,
and this translates into the range of $M_c$ for a given $N_l$.  Taking 
$M_* R \simeq 100$, we find that the compactification scale must be 
in the range
\begin{eqnarray}
  5 \times 10^{14}~{\rm GeV} 
  \;\lsim\; M_c \;\lsim\; 
  2 \times 10^{15}~{\rm GeV},
\label{eq:mc}
\end{eqnarray}
which is considerably lower than the usual 4d unification scale 
$M_U \simeq 2 \times 10^{16}~{\rm GeV}$.  Since the mass for the broken 
gauge bosons is given by $1/R$, it induces the $d=6$ proton decay at a 
rate contradicting the bound from Super-Kamiokande \cite{Shiozawa:1998si}, 
if quarks and leptons are localized on the $SU(5)$-preserving brane.  
In fact, this constraint was used in Ref.~\cite{Contino:2001si} to 
conclude that strict NDA assumption does not work, and the contribution 
from unknown ultraviolet physics is needed to obtain the right values 
for $\sin^2\theta_w$ and $\alpha_s$.  In other words, the contribution 
from KK towers alone is insufficient to explain the small difference 
of $\sin^2\theta_w$ between the experiment and naive 4d GUT prediction, 
since $M_* R$ must be smaller than $\sim 10$ from the proton decay 
constraint.\footnote{
The constraint from $d=6$ proton decay was also used in 
Ref.~\cite{Cheng:2001qp} to conclude that $N_l$ ($\simeq M_* R$) must be 
smaller than $\sim 20$ and that the calculable contribution from KK 
towers cannot explain the discrepancy of the gauge coupling values 
between the experiment and the theoretical prediction of 4d GUT.
Ref.~\cite{Csaki:2001qm} also argues that $N_l$ must be smaller than 
$\sim 25$ using $\alpha N_l \lsim 1$ to estimate the strong coupling 
bound, while we here use $\alpha N_l/4 \pi \lsim 1$ to estimate it.}
In the present case, however, the first generation matter lives in 
the bulk so that the constraint from $d=6$ proton decay is evaded even 
if the compactification scale is low.  This allows us to consider larger 
values for $M_* R$, that is, the scenario where the theory is strongly 
coupled at the cut-off scale.  Then, the calculable contribution from 
KK towers could completely explain the small discrepancy of 
$\sin^2\theta_w$ between the experimental and theoretical values that 
was present in the case of the minimal 4d GUT with a single threshold.  
Note that the non-calculable contribution from unknown physics above 
$M_*$ is expected to be small in this case through NDA in higher 
dimensions.

We here consider uncertainties for the present analysis.  Since the 
theory is assumed to be strongly coupled at the cut-off scale, higher 
order effects could be important around that energy scale.  However, 
the logarithmic contribution from KK towers discussed above comes from 
entire energy range from $1/R$ to $M_*$, and the theory is weakly 
coupled in most of this energy region.  Actually, various interactions 
quickly become weak below $M_*$, suppressed by powers of $(E/M_*)$ 
at energy scale $E$.  This is because in the 5d picture the couplings 
in the theory have negative mass dimensions, and in the 4d picture 
the number of KK states circulating in the loop decreases with 
decreasing energies so that loop expansion parameters in the 
theory ('t Hooft couplings) become small by powers of $(E/M_*)$.  
Therefore, we expect that the the leading log calculation of the 
threshold correction to $\sin^2\theta_w$ is reliable at least 
at the order of magnitude level, although the precise coefficients may 
receive corrections from this higher order effect.  To be more precise, 
the difference of the gauge couplings runs logarithmically in all energy 
regions between $1/R$ and $M_*$, and the one-loop estimates are reliable 
only when the renormalization scale is at least a factor of a few smaller 
than $M_*$; higher loop effects would equally be important around 
the cut-off scale.  This would give $O(10\%)$ uncertainties in the 
calculations of the {\it threshold corrections} of $\sin^2\theta_w$ and 
$\alpha_s$.  A similar size of uncertainties is also expected from 
tree-level $SU(5)$-breaking boundary operators.  We emphasize that 
the uncertainties are for the threshold corrections and are not 
$O(10\%)$ uncertainties for the values of $\sin^2\theta_w$ and 
$\alpha_s$ themselves.

We then find that the observed values for the gauge coupling constants 
are well reproduced by taking $M_* R = O(10^2 - 10^3)$.  That is, we can 
explain the difference of $\sin^2\theta_w$ (and $\alpha_s$) between the 
experimental value and the theoretical prediction obtained by assuming 
the exact gauge unification at a single threshold.  We note that the 
situation is better in the present scenario than in usual 
4d GUTs.\footnote{
We thank Lawrence Hall for stressing this point.}
Let us consider, for example, predicting $\alpha_s$ from the observed 
values of $\sin^2\theta_w$ and $e$.  It is known that if we calculate 
$\alpha_s$ without including any threshold correction, we obtain a somewhat 
larger value $\alpha_s|_{\rm th} \simeq 0.130$ \cite{Langacker:1993rq} 
than the experimentally measured value $\alpha_s|_{\rm ex} \simeq 
0.118 \pm 0.002$ \cite{Groom:2000in}.  Thus, we have to explain the 
difference $\alpha_s|_{\rm ex} - \alpha_s|_{\rm th} \simeq -0.012 \pm 0.002$ 
by the GUT-scale threshold correction $\Delta_{\alpha_s}^{\rm gut}$.  
(Here we ignore the weak-scale threshold corrections, which typically 
give $|\Delta_{\alpha_s}^{\rm weak}| \lsim 0.004$.)  In usual 4d GUTs, 
the size of the GUT-scale threshold correction is given by 
$|\Delta_{\alpha_s}^{\rm gut}| \lsim 0.02$, but we cannot predict the 
value of $\Delta_{\alpha_s}^{\rm gut}$ in general since it strongly 
depends on the mass spectrum of the GUT scale particles.\footnote{
In the minimal supersymmetric $SU(5)$ GUT in 4d, 
$\Delta_{\alpha_s}^{\rm gut}$ is positive in most of the parameter 
space, due to the large mass for the triplet Higgses required to 
satisfy the bound from $d=5$ proton decay.}
On the other hand, in the present case, we completely know the 
pattern of the GUT-scale particle masses, so that we can calculate the 
threshold correction, $\Delta_{\alpha_s}^{\rm gut}$, for a given value 
of $M_* R$.  It is given by 
$\Delta_{\alpha_s}^{\rm gut} \simeq -(3/7\pi) \alpha_s^2 \ln(M_* R)$.  
Numerically, we find $\Delta_{\alpha_s}^{\rm gut} \simeq -0.009 \pm 0.002$ 
($\Delta_{\alpha_s}^{\rm gut} \simeq -0.013 \pm 0.003$) if $M_* R = 100$ 
($M_* R = 1000$), where the errors represent the $O(10\%)$ uncertainties 
discussed before.  We find that the observed value of $\alpha_s$ is well 
reproduced with the values of $M_* R$ suggested by the NDA analysis.

Of course, we cannot prove that these values of $M_* R$ exactly give 
a {\it truly} strongly coupled theory at the cut-off scale (all $\eta$'s 
equal to 1), since there are many uncertainties in estimating the 
over-all value for the gauge coupling (but not the differences between 
the three couplings) at the cut-off scale.  For example, the contribution 
from $SU(5)$ symmetric power-law running (scheme dependence, in other 
words) could change the value.  However, within the uncertainties in 
estimating various quantities, we can say that the scenario where the 
theory is (moderately) strongly coupled at the cut-off scale is 
consistent with low-energy observations.  It is particularly interesting 
that the $M_* R$ value giving the desired low-energy gauge coupling 
values is consistent with the requirement that the theory is strongly 
coupled at the cut-off scale.

\section{Other Issues}
\label{sec:other}

In the model discussed in the previous sections, $d=6$ proton decay 
occurs through mixings between the first and heavier generations 
occurring at the coupling to the heavy broken gauge bosons.\footnote{
Similar situations are also discussed in the context of 
dimensionally deconstructed models \cite{Csaki:2001qm}.}  
Thus, their rates are suppressed by mixing angles that are expected to 
have similar order of magnitudes to the corresponding CKM angles.  In 
the present case where ${\bf 10}_2$ resides in the bulk and ${\bf 5}^*_2$ 
lives on the brane, the dominant decay mode is $K^+ \nu_\mu$ or 
$\mu^+ \pi^0$.  However, their amplitudes receive suppression of order 
$V_{ub} V_{cb} V_{e3}$ and $V_{ub}^2 V_{e2}$, respectively, which are 
$10^{-5} - 10^{-6}$.  Therefore, the lifetime is roughly 
$10^{40}~{\rm years}$, and there would be little hope for detection in 
the near future.  In the case where both ${\bf 10}_2$ and ${\bf 5}^*_2$ 
are in the bulk, the dominant decay mode is $K^+ \nu_\tau$, whose 
amplitude also receives suppression of order $V_{ub} V_{cb} V_{e3} 
\sim 10^{-5}$.  On the other hand, if ${\bf 10}_2$ is on the brane and 
${\bf 5}^*_2$ is in the bulk, the $p \rightarrow K^+ \nu_\tau$ and 
$p \rightarrow \mu^+ K^0$ decays could occur with only $V_{us} V_{e3} 
\sim V_{us}^2 \sim 10^{-2}$ suppression in their amplitudes, that is, 
with the lifetime of $10^{33} - 10^{35}~{\rm years}$.  Incidentally, 
if $M_* R$ is somewhat larger than 100, the proton lifetime becomes 
shorter.  In the case of $M_* R \simeq 500$ ($M_* R \simeq 1000$), for 
example, the lifetime becomes factor 70 ($450$) shorter compared with 
the case of $M_* R \simeq 100$.  Thus, in the case where ${\bf 10}_2$ 
is on the brane, it is probable that strange $d=6$ proton decay, 
involving the second generation particles in the final state, could be 
discovered in future experiments.

Finally, we estimate 4d reduced Planck scale $M_P$ assuming that the 
strength of the gravitational interaction is also dictated by the NDA 
analysis.  Since the theory is strongly coupled at $M_*$, the kinetic term 
for the graviton is given by $S = \int d^4x dy (M_*^3/16 \pi^3) {\cal R}$, 
where ${\cal R}$ is the Ricci scalar.  Thus, after integrating $y$, $M_P$ 
is given by $M_P^2 = M_*^3 R / (16 \pi^2)$.  Substituting the value 
obtained in Eq.~(\ref{eq:mc}) with $M_* R \simeq 100$, we obtain 
$M_P \simeq 10^{17}~{\rm GeV}$.  This is substantially higher than the 
4d unification scale $M_U \simeq 2 \times 10^{16}~{\rm GeV}$, but still 
somewhat lower than the observed value $M_P \simeq 2 \times 
10^{18}~{\rm GeV}$.  To reproduce the observed value, we need either 
an $O(10)$ coefficient, $M_* R \gsim 1000$, or $n$ extra dimensions with 
radius $R \simeq O(10^{2/n})$ in which (only) gravity propagates.
However, it is true that $M_P$ is an order of magnitude separated from 
the apparent unification scale $M_U$ by the presence of the large extra 
dimension necessary to break the GUT symmetry.  The precise estimate is 
also dependent on the number of extra dimensions, gauge group and matter 
content, which we here took those of the minimal 5d $SU(5)$ model as 
a representative case.  Thus, we expect that the existence of this type 
of dimension may provide a general way of separating the two scales 
in string theory.

\section{Conclusions}
\label{sec:concl}

In this paper, we have explored the scenario where all the couplings in 
the theory are strong at the cut-off scale, in the context of higher 
dimensional grand unified field theories.  This provides a calculable 
framework for gauge coupling unification in higher dimensions.  The 
non-calculable effect from unknown ultraviolet physics is suppressed by 
assuming that all the operators in the theory scale according to naive 
dimensional analysis in higher dimensions \cite{Hall:2001pg}.  Then, the 
threshold correction to $\sin^2\theta_w$ dominantly comes from the 
calculable contribution from KK towers, giving the values for 
$\sin^2\theta_w$ and $\alpha_s$ in good agreement with low-energy data.  
Although the theory is strongly coupled at the cut-off scale $M_*$, it 
quickly becomes weakly coupled below $M_*$, allowing reliable estimates 
of threshold corrections to the gauge coupling unification.  A crucial 
point is that we can have large values of $M_* R$ without conflicting 
with the constraint from proton decay by putting the first generation 
matter in the bulk.  This enables us to consider the strong coupling 
scenario, in contrast with the previous work \cite{Contino:2001si} where 
it was concluded that $M_* R$ must be smaller than $\sim 10$ due to the 
proton decay constraint and thus the calculable contribution from KK 
towers cannot fully explain the low-energy data.

We have shown that the ansatz where all the coupling constants are 
dictated by naive dimensional analysis in higher dimensions is consistent 
with low-energy observations.  We have presented a completely realistic 
5d supersymmetric $SU(5)$ model as an explicit example.  This suggests 
that the higher dimensional grand unified theory is a low-energy effective 
theory of some more fundamental theory that is strongly coupled at the 
scale $M_*$.  In the present scenario, the observed weakness of various 
couplings is attributed to the presence of a moderately large extra 
dimension(s).  The hierarchy among various couplings arise from different 
numbers of dimensions in which various fields propagate.  The presence of 
this large dimension(s) is required to solve the problems in conventional 
GUTs, such as doublet-triplet splitting and $d=5$ proton decay problems, 
by extra dimensional mechanisms while preserving successful prediction of 
$\sin^2\theta_w$ \cite{Hall:2001pg, Kawamura:2001ev}.  Therefore, in this 
framework, solving the many conventional problems in GUTs is transformed 
to finding a single mechanism of naturally getting such a large extra 
dimension(s) with the radius of order $10^2 - 10^3$ in units of the 
fundamental scale.  It would be interesting to consider a mechanism of 
generating this type of large extra dimension(s) in the context of 
string theory.

\vspace{5mm}

{\bf Note added}

While this work is being completed, we received Ref.~\cite{Watari:2001pj}
where it is hoped that $SU(5)$-breaking boundary operators may not exist 
when the gauge group is broken only by orbifold reflections (not 
translations).  Even then, however, there are fixed points which do not 
preserve full $SU(5)$ symmetry.  Thus, $SU(5)$-violating local operators 
can be written on these points, since they are not prohibited by the 
restricted gauge symmetry.

\section*{Acknowledgements}

The author would like to thank L.~Hall for reading the manuscript and 
useful discussions.  He also thanks K.-I.~Izawa, T.~Kugo, T.~Moroi, 
H.~Nakano, T.~Okui, M.~Yamaguchi and T.~Yanagida for valuable discussions, 
and the Summer Institute 2001 at Yamanashi, Japan, for a kind hospitality.  
This work was supported by the Miller Institute for Basic Research 
in Science and the Director, Office of Science, Office of High Energy 
and Nuclear Physics, of the U.S. Department of Energy under Contract 
DE-AC03-76SF00098.


\begin{thebibliography}{99}

\bibitem{Georgi:1974yf}
H.~Georgi, H.~R.~Quinn and S.~Weinberg,
Phys.\ Rev.\ Lett.\ {\bf 33}, 451 (1974); \\
S.~Dimopoulos, S.~Raby and F.~Wilczek,
Phys.\ Rev.\ D {\bf 24}, 1681 (1981);\\
L.~E.~Ibanez and G.~G.~Ross,
Phys.\ Lett.\ B {\bf 105}, 439 (1981).

\bibitem{Georgi:1974sy}
H.~Georgi and S.~L.~Glashow,
Phys.\ Rev.\ Lett.\ {\bf 32}, 438 (1974); \\
S.~Dimopoulos and H.~Georgi,
Nucl.\ Phys.\ B {\bf 193}, 150 (1981); \\
N.~Sakai,
Z.\ Phys.\ C {\bf 11}, 153 (1981).

\bibitem{Sakai:1982pk}
N.~Sakai and T.~Yanagida,
Nucl.\ Phys.\ B {\bf 197}, 533 (1982); \\
S.~Weinberg,
Phys.\ Rev.\ D {\bf 26}, 287 (1982).

\bibitem{Candelas:1985en}
P.~Candelas, G.~T.~Horowitz, A.~Strominger and E.~Witten,
Nucl.\ Phys.\ B {\bf 258}, 46 (1985);\\
E.~Witten,
Nucl.\ Phys.\ B {\bf 258}, 75 (1985);\\
J.~D.~Breit, B.~A.~Ovrut and G.~C.~Segre,
Phys.\ Lett.\ B {\bf 158}, 33 (1985);\\
A.~Sen,
Phys.\ Rev.\ Lett.\  {\bf 55}, 33 (1985).

\bibitem{Dixon:jw}
L.~J.~Dixon, J.~A.~Harvey, C.~Vafa and E.~Witten,
Nucl.\ Phys.\ B {\bf 261}, 678 (1985);
Nucl.\ Phys.\ B {\bf 274}, 285 (1986);\\
L.~E.~Ibanez, J.~E.~Kim, H.~P.~Nilles and F.~Quevedo,
Phys.\ Lett.\ B {\bf 191}, 282 (1987).

\bibitem{Ibanez:1991zv}
See, for example, 
L.~E.~Ibanez, D.~Lust and G.~G.~Ross,
Phys.\ Lett.\ B {\bf 272}, 251 (1991)
[hep-th/9109053].

\bibitem{Hall:2001pg}
L.~Hall and Y.~Nomura,
Phys.\ Rev.\ D {\bf 64}, 055003 (2001)
[hep-ph/0103125].

\bibitem{Kawamura:2001ev}
Y.~Kawamura,
Prog.\ Theor.\ Phys.\  {\bf 105}, 999 (2001)
[hep-ph/0012125].

\bibitem{Barbieri:2001vh}
R.~Barbieri, L.~J.~Hall and Y.~Nomura,
Phys.\ Rev.\ D {\bf 63}, 105007 (2001)
[hep-ph/0011311].

\bibitem{Altarelli:2001qj}
G.~Altarelli and F.~Feruglio,
Phys.\ Lett.\ B {\bf 511}, 257 (2001)
[hep-ph/0102301].

\bibitem{Hebecker:2001wq}
A.~Hebecker and J.~March-Russell,
hep-ph/0106166.

\bibitem{Barbieri:2001yz}
R.~Barbieri, L.~J.~Hall and Y.~Nomura,
hep-ph/0106190.

\bibitem{Kawamura:2001ir}
Y.~Kawamura,
Prog.\ Theor.\ Phys.\  {\bf 105}, 691 (2001)
[hep-ph/0012352];\\
A.~B.~Kobakhidze,
Phys.\ Lett.\ B {\bf 514}, 131 (2001)
[hep-ph/0102323];\\
R.~Barbieri, L.~J.~Hall and Y.~Nomura,
hep-th/0107004;\\
A.~Hebecker and J.~March-Russell,
hep-ph/0107039;\\
T.~Li,
hep-th/0107136;
hep-ph/0108120;\\
J.~A.~Bagger, F.~Feruglio and F.~Zwirner,
hep-th/0107128;\\
A.~Masiero, C.~A.~Scrucca, M.~Serone and L.~Silvestrini,
hep-ph/0107201;\\
N.~Haba, Y.~Shimizu, T.~Suzuki and K.~Ukai,
hep-ph/0107190;\\
L.~J.~Hall, H.~Murayama and Y.~Nomura,
hep-th/0107245;\\
N.~Haba, T.~Kondo, Y.~Shimizu, T.~Suzuki and K.~Ukai,
hep-ph/0108003;\\
T.~Asaka, W.~Buchmuller and L.~Covi,
hep-ph/0108021;\\
L.~Hall, Y.~Nomura, T.~Okui and D.~Smith,
hep-ph/0108071;\\
R.~Dermisek and A.~Mafi,
hep-ph/0108139.

\bibitem{Hall:2001zb}
L.~Hall, Y.~Nomura and D.~Smith,
hep-ph/0107331.

\bibitem{Hall:2001rz}
L.~Hall, J.~March-Russell, T.~Okui and D.~Smith,
hep-ph/0108161.

\bibitem{Nomura:2001mf}
Y.~Nomura, D.~Smith and N.~Weiner,
Nucl.\ Phys.\ B {\bf 613}, 147 (2001)
[hep-ph/0104041].

\bibitem{Csaki:2001qm}
C.~Csaki, G.~D.~Kribs and J.~Terning,
hep-ph/0107266.

\bibitem{Cheng:2001qp}
H.~Cheng, K.~T.~Matchev and J.~Wang,
hep-ph/0107268.

\bibitem{Contino:2001si}
R.~Contino, L.~Pilo, R.~Rattazzi and E.~Trincherini,
hep-ph/0108102.

\bibitem{Chacko:2000hg}
Z.~Chacko, M.~A.~Luty and E.~Ponton,
JHEP{\bf 0007}, 036 (2000)
[hep-ph/9909248].

\bibitem{Chanowitz:1977ye}
M.~S.~Chanowitz, J.~Ellis and M.~K.~Gaillard,
Nucl.\ Phys.\ B {\bf 128}, 506 (1977).

\bibitem{Fukuda:1998mi}
Y.~Fukuda {\it et al.}  [Super-Kamiokande Collaboration],
Phys.\ Rev.\ Lett.\  {\bf 81}, 1562 (1998)
[hep-ex/9807003].

\bibitem{Kaplan:2000ac}
D.~E.~Kaplan, G.~D.~Kribs and M.~Schmaltz,
Phys.\ Rev.\ D {\bf 62}, 035010 (2000)
[hep-ph/9911293]; \\
Z.~Chacko, M.~A.~Luty, A.~E.~Nelson and E.~Ponton,
JHEP{\bf 0001}, 003 (2000)
[hep-ph/9911323].

\bibitem{Pierce:1997zz}
See, for example,
D.~M.~Pierce, J.~A.~Bagger, K.~T.~Matchev and R.~Zhang,
Nucl.\ Phys.\ B {\bf 491}, 3 (1997)
[hep-ph/9606211].

\bibitem{Seesaw}
T.~Yanagida, 
in {\it Proc. of the Workshop on the Unified Theory and 
Baryon Number in the Universe}, 
ed. O.~Sawada and A.~Sugamoto 
(KEK report 79-18, 1979), p. 95; \\
M.~Gell-Mann, P.~Ramond, and R.~Slansky, 
in {\it Supergravity}, 
ed. P.~van Nieuwenhuizen and D.Z.~Freedman 
(North Holland, Amsterdam, 1979), p. 315.

\bibitem{Dienes:1998vh}
K.~R.~Dienes, E.~Dudas and T.~Gherghetta,
Phys.\ Lett.\ B {\bf 436}, 55 (1998)
[hep-ph/9803466];
Nucl.\ Phys.\ B {\bf 537}, 47 (1999)
[hep-ph/9806292].

\bibitem{Hisano:1992mh}
J.~Hisano, H.~Murayama and T.~Yanagida,
Phys.\ Rev.\ Lett.\ {\bf 69}, 1014 (1992).

\bibitem{Shiozawa:1998si}
M.~Shiozawa {\it et al.}  [Super-Kamiokande Collaboration],
Phys.\ Rev.\ Lett.\ {\bf 81}, 3319 (1998)
[hep-ex/9806014].

\bibitem{Langacker:1993rq}
See, for example,
P.~Langacker and N.~Polonsky,
Phys.\ Rev.\ D {\bf 47}, 4028 (1993)
[hep-ph/9210235];
Phys.\ Rev.\ D {\bf 52}, 3081 (1995)
[hep-ph/9503214].

\bibitem{Groom:2000in}
D.~E.~Groom {\it et al.}  [Particle Data Group Collaboration],
Eur.\ Phys.\ J.\ C {\bf 15}, 1 (2000).

\bibitem{Watari:2001pj}
T.~Watari and T.~Yanagida,
hep-ph/0108152.

\end{thebibliography}
\end{document}